\documentclass[aps,prl,twocolumn,superscriptaddress,showpacs,draft]{revtex4}
\usepackage{graphicx}
\input{epsf}

\begin{document}

\title{Google matrix, dynamical attractors and Ulam networks}

\author{D.L.Shepelyansky}
%\homepage[]{http://www.quantware.ups-tlse.fr}
\affiliation{\mbox{Laboratoire de Physique Th\'eorique (IRSAMC), 
Universit\'e de Toulouse, UPS, F-31062 Toulouse, France}}
\affiliation{\mbox{LPT (IRSAMC), CNRS, F-31062 Toulouse, France}}
\author{O.V.Zhirov}
\affiliation{\mbox{Budker Institute of Nuclear Physics, 
630090 Novosibirsk, Russia}}
\affiliation{\mbox{LPT (IRSAMC), CNRS, F-31062 Toulouse, France}}

%\date{\today}
\date{May  26, 2009; Revised: August 20, 2009 }

%\PACS{
%{05.45.Ac}{Low-dimensional chaos}
%\and
% 89.20.Hh  World Wide Web, Internet
%\and
%05.45.-a Nonlinear dynamics and chaos}

\begin{abstract}
We study the properties of the Google matrix
generated by a coarse-grained Perron-Frobenius operator of
the Chirikov typical map with dissipation. The finite size matrix 
approximant of this operator is constructed by the Ulam method.
This method applied to the simple dynamical model 
creates the directed Ulam networks
with approximate scale-free scaling and 
characteristics being rather similar to those of 
the  World Wide Web. The simple dynamical attractors play here the role
of popular web sites with a strong concentration of PageRank.
A variation of the Google parameter $\alpha$ or other 
parameters of the dynamical map can drive 
the PageRank of the Google matrix to a delocalized phase 
with a strange attractor where the Google search becomes inefficient.
\end{abstract}

\maketitle

\section{I Introduction}

 The World Wide Web (WWW) continues its striking 
expansion going beyond $10^{11}$ web pages. Information retrieval
from such an enormous database becomes the main challenge
for WWW users. An efficient solution, known as the PageRank Algorithm (PRA)
proposed by Brin and Page in 1998 \cite{brin}, forms the basis of 
the Google search engine used by the majority of internautes in everyday life.
The PRA is based on the construction of the Google matrix
which can be written as (see e.g. \cite{googlebook} for details):
\begin{equation}
{\bf G}=\alpha {\bf S}+(1-\alpha) {\bf E}/N \; .
\label{eq1}
\end{equation}
Here the matrix ${\bf S}$ is constructed from the adjacency matrix  ${\bf A}$
of  directed network links between $N$ nodes 
so that $S_{ij}=A_{ij}/\sum_k A_{kj}$ and
the elements of columns with
only zero elements are replaced by $1/N$. The  second term
in r.h.s. of (\ref{eq1}) describes  a finite probability $1-\alpha$
for WWW surfer to jump at random to any node so that the matrix elements
$E_{ij}=1$. This term allows to stabilize the convergence of PRA
introducing a gap between the maximal eigenvalue $\lambda=1$
and other eigenvalues $\lambda_i$. Usually the Google search uses the value
$\alpha=0.85$ \cite{googlebook}. By the construction 
$\sum_i G_{ij}=1$ so that the asymmetric matrix ${\bf G}$
has a left eigenvector being a homogeneous constant for $\lambda=1$.
The right eigenvector at $\lambda=1$ is the PageRank vector with 
positive elements $p_j$ and $\sum_j p_j=1$. All WWW nodes can be
ordered by decreasing $p_j$ so that the PageRank plays a primary role in the
ordering of websites and information retrieval.
The classification of nodes in the decreasing order of
$p_j$ values is used by the Google search to classify
importance of web nodes. The information retrieval
and ordering is based on this classification and we also use it in the
following.

It is interesting and important to note that 
by the construction the operator ${\bf G}$ belongs to the
class of Perron-Frobenius operators \cite{googlebook}.
Such type of operators naturally appear in the ergodic theory
\cite{sinai}
and in the description of dynamical systems
with Hamiltonian or dissipative dynamics \cite{mbrin,osipenko}. 

The studies of properties of  ${\bf G}$ are usually 
done only for the PageRank vector which can be find efficiently by
the PRA due to a relatively small average number of links in WWW.
At present Google succeeds to operate with PageRank vectors
of size of the whole WWW being of the order of $10^{11}$.
It is established that for large WWW subsets $p_j$
is satisfactory described by a scale-free algebraic decay
with $p_j \sim 1/j^{\beta}$ where $j$ is the PageRank ordering index
and $\beta \approx 0.9$ \cite{googlebook,donato}.
The studies of PageRank properties are now very active 
in the computer science community being presented in a number
of interesting publications (see e.g. \cite{boldi,avrach1,avrach2}
and an overview of the field in \cite{avrach3}).

While the properties of the PageRank are of primary
importance it is also interesting to analyze the properties
of the Google matrix ${\bf G}$ as a whole large matrix.
Such an analysis can help to establish links between
the Google matrix and other fields of physics where
large matrices play an important role.
Among such fields we can mention the Random matrix theory
\cite{mehta} which finds applications for a description of spectra
in complex many-body quantum systems and the Anderson localization 
which is an important physical phenomenon for an electron transport
in disordered systems (see e.g. \cite{anderson}).
A transition from localized to delocalized eigenstates also
can take place in networks of small world type
(see \cite{giraud2005,berkovits}). However, in the physical
systems considered in \cite{mehta,anderson,giraud2005,berkovits}
all matrices are Hermitian with real eigenvalues
while the Perron-Frobenius matrices have 
generally complex eigenvalues.

A first attempt to analyze the properties of
right eigenvectors $\psi_i$ (${\bf G} \psi_i=\lambda_i \psi_i$)
and complex eigenvalues $\lambda_i$ was done recently in \cite{ggs}.
The Google matrix was constructed from a directed network
generated by the Albert-Barabasi model and the WWW University
networks with randomization of links. The Google matrix was
considered mainly for the value $\alpha=0.85$. It was shown that
at certain conditions a delocalization phase emerges
for the PageRank and states with complex $\lambda$. In spite of a number
of interesting results found in \cite{ggs} a weak feature
of models used there is a significant gap between
$\lambda=1$ of PageRank vector and $|\lambda_i| \leq 0.4$ of other vectors.
We note that according to \cite{ggs} the University networks
have $|\lambda_i|$ close to $1$ but after randomization of links
a large gap emerges in the spectrum of $\lambda$.
This gap in $|\lambda|$ was rather large and was not sensitive
to a variation of $\alpha$ in the interval $0.85 \leq \alpha \leq 1$.
Hence, the PageRank vector also was not very sensitive to $\alpha$
while  for real WWW 
it is know that $p_j$ is rather sensitive to $\alpha$
due to existence of $|\lambda_i|$ close to $1$ \cite{googlebook,ggs}. 
Thus the results obtained in \cite{ggs} show that even if 
the Google matrix is constructed on the basis of typical models
of scale-free networks it is quite possible that
its spectrum may have a large gap
for $0.85 \leq \alpha \leq 1$ thus being rather far from 
spectral properties of the Google matrices of WWW.
Therefore it is rather desirable to have other simple models
which generate a directed network with Google matrix 
properties being close to those of WWW.
%\vskip 0.3cm
\begin{figure}
%\centerline{\epsfxsize=8.5cm\epsffile{fig1prl.eps}}
\centerline{\epsfxsize=2.7cm\epsffile{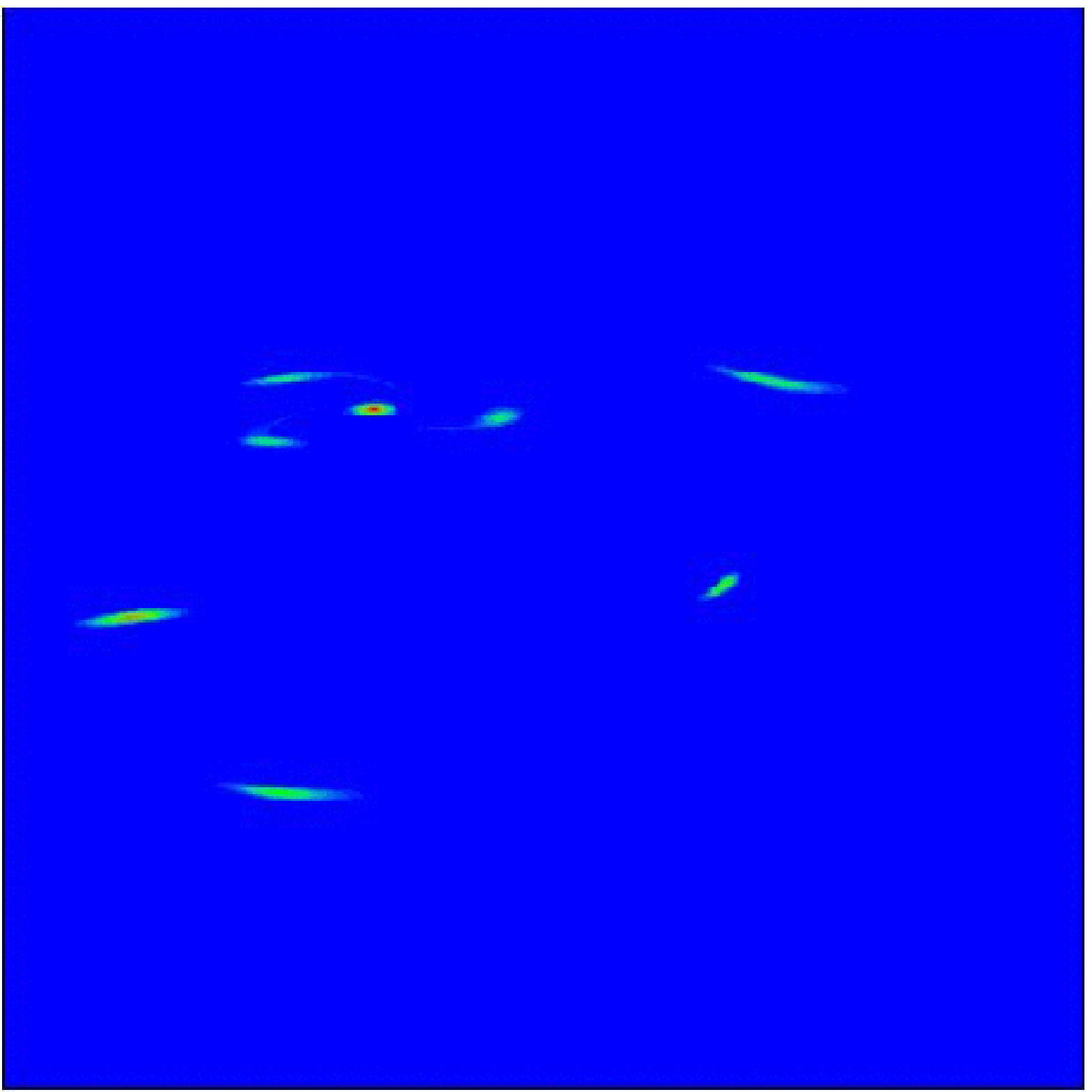}
\hfill\epsfxsize=2.7cm\epsffile{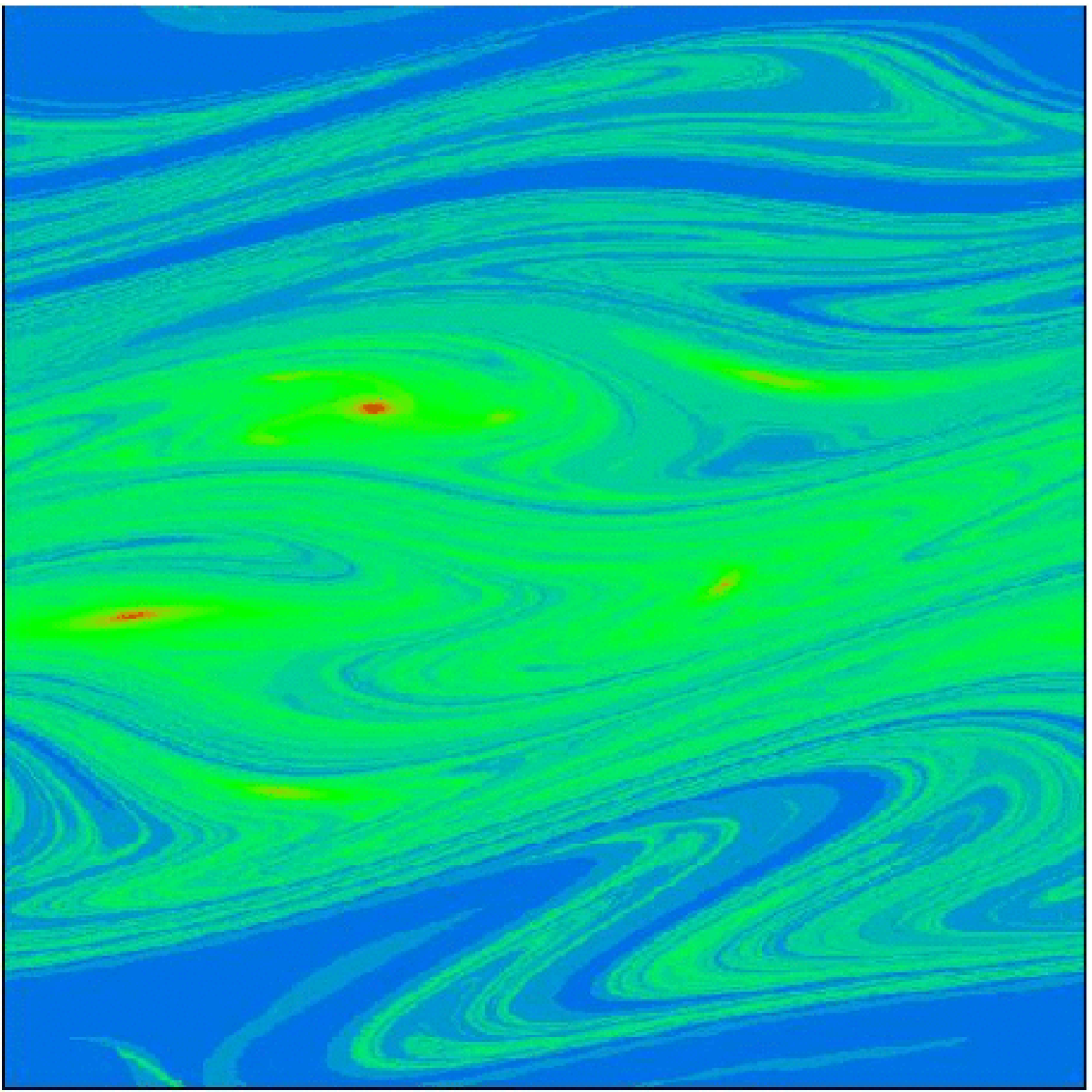}
\hfill\epsfxsize=2.7cm\epsffile{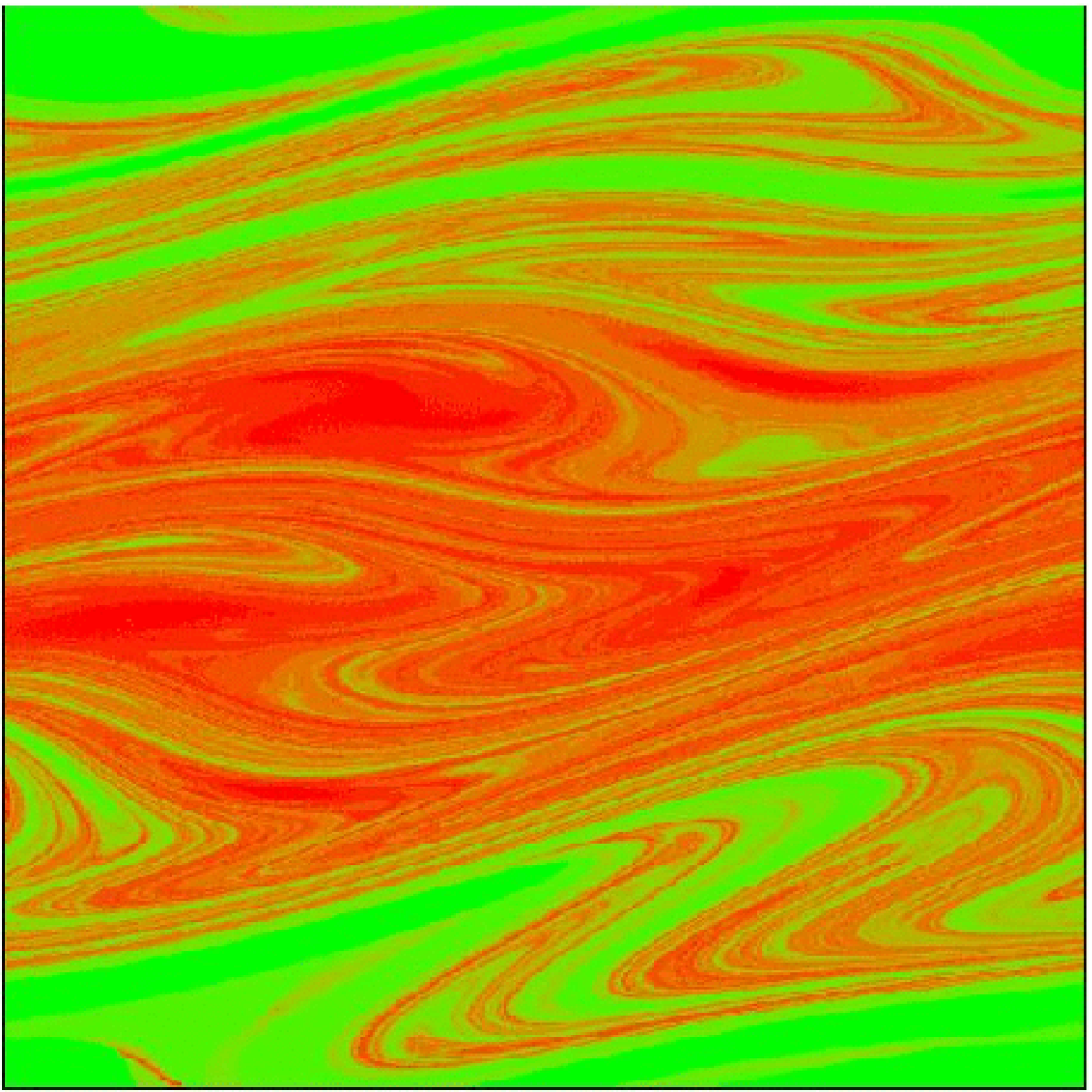}}
\vglue 0.2cm
\centerline{\epsfxsize=2.7cm\epsffile{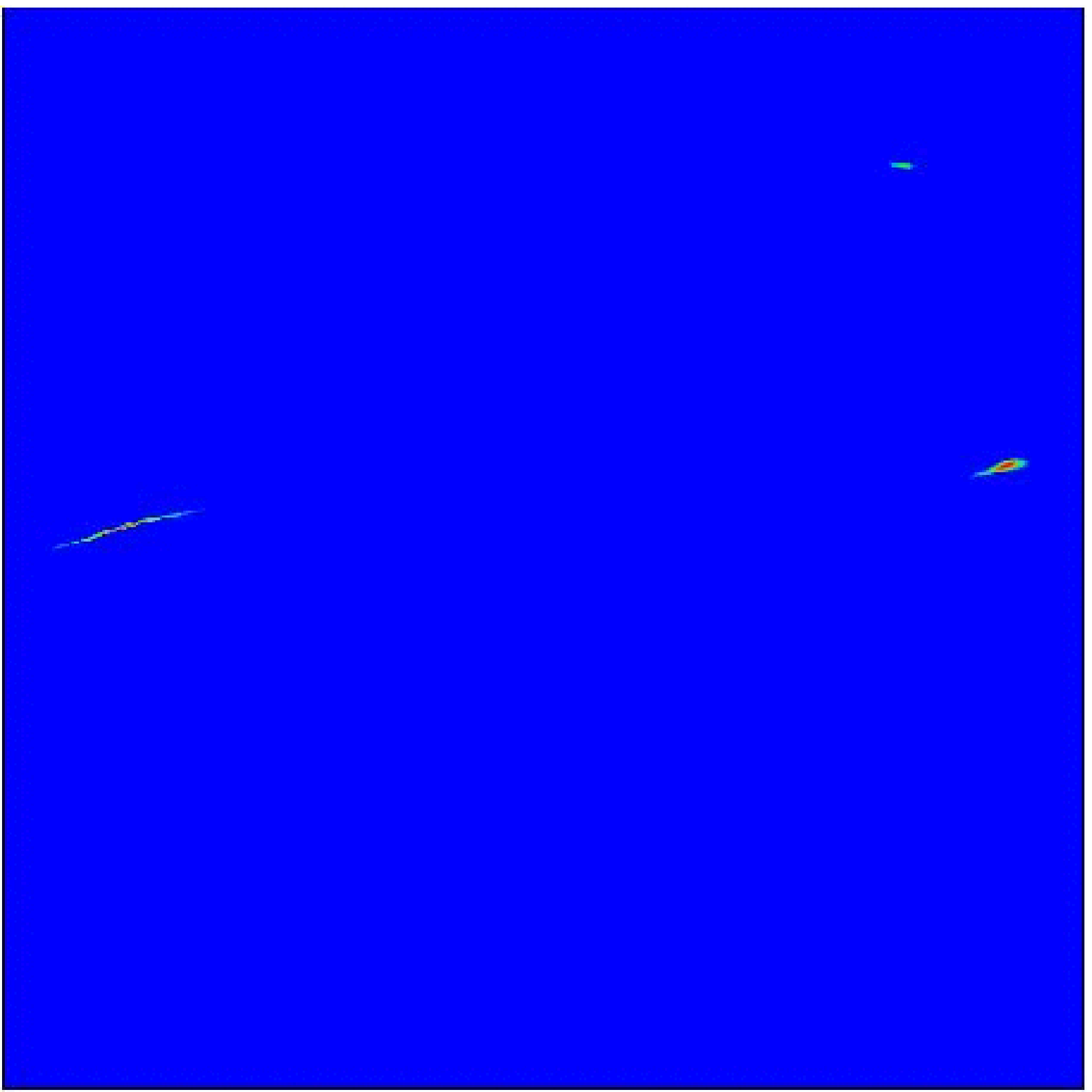}
\hfill\epsfxsize=2.7cm\epsffile{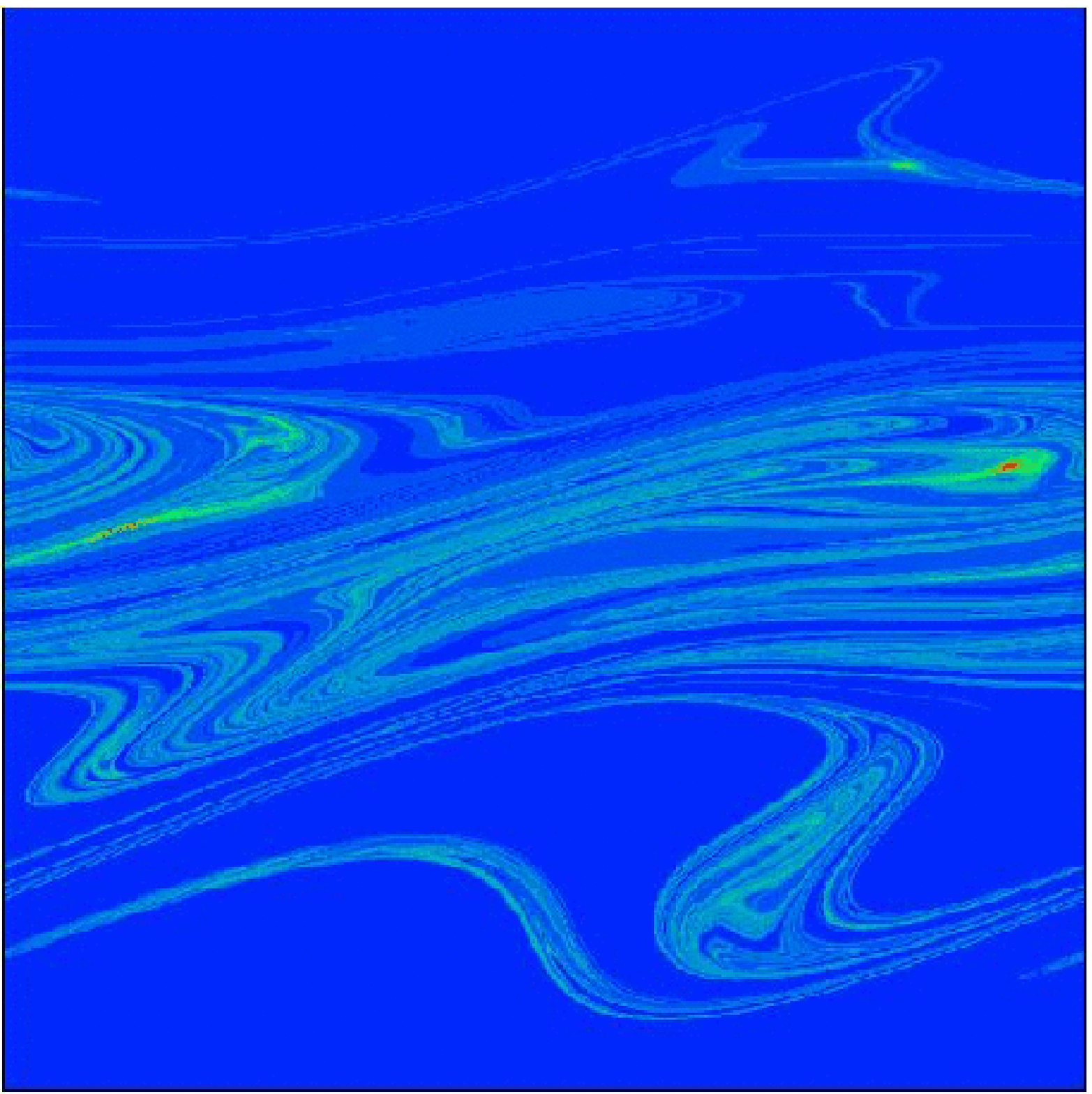}
\hfill\epsfxsize=2.7cm\epsffile{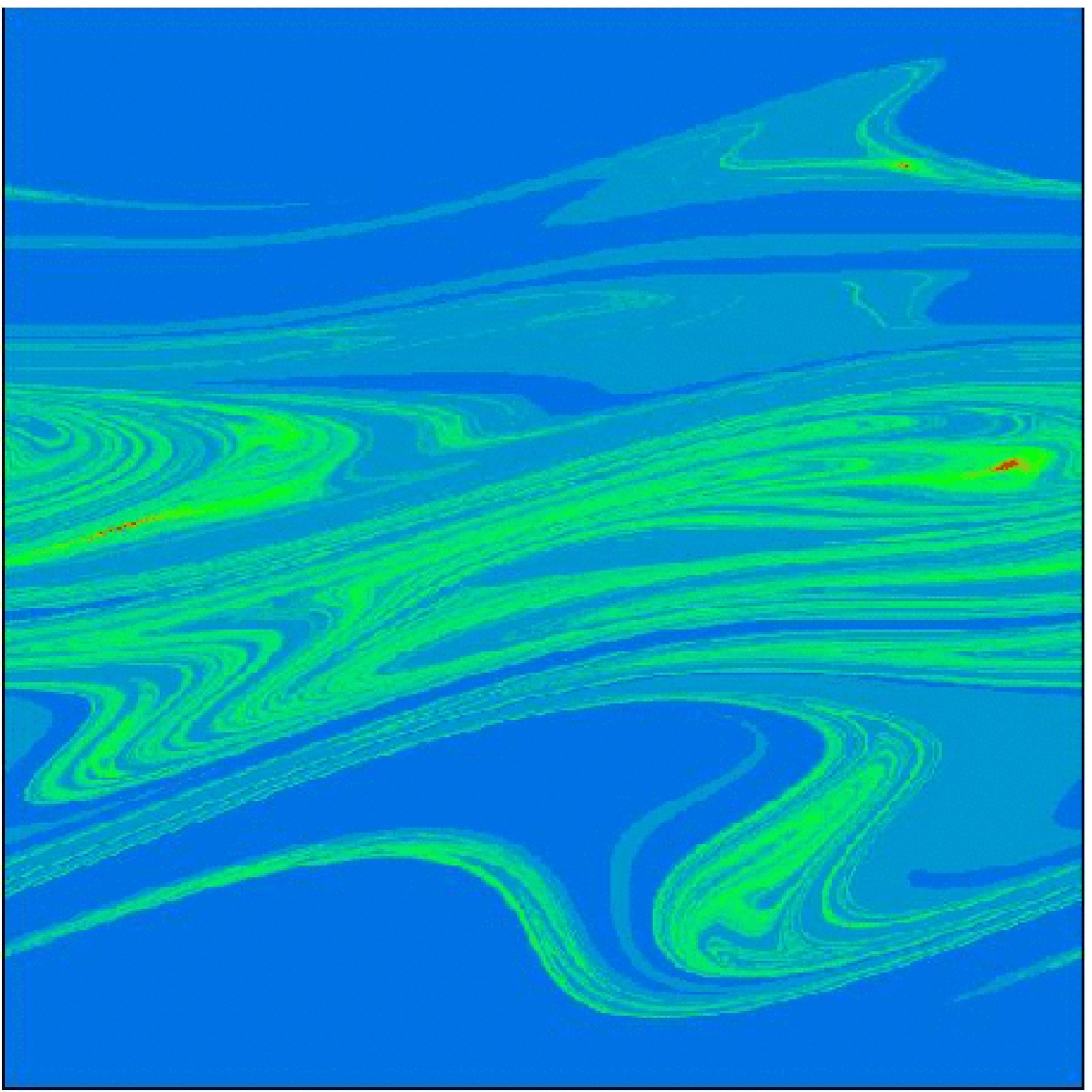}}
\vglue -0.2cm
\caption{(Color online) PageRank $p_j$ for the Google matrix generated by
the Chirikov typical map (\ref{eq2}) at
$T=10$, $k=0.22$, $\eta=0.99$ (set $T10$, top row)
and  $T=20$, $k=0.3$, $\eta=0.97$ (set $T20$, bottom row)
with  $\alpha=1, 0.95, 0.85$ (left to right).
The phase space region  $0 \leq x < 2\pi; -\pi \leq p < \pi $ is divided on
$N=3.6 \cdot 10^5$ cells; $p_j$
is zero for blue and maximal for red.
} 
\label{fig1}
\end{figure}

With an aim to have more realistic models
we develop in this work another approach and construct the Google matrix from
the Perron-Frobenius operator generated by
a certain dynamical system. 
The probability flow in these models has rich and nontrivial features
of general importance like simple and strange attractors
with localized and delocalized dynamics governed by simple dynamical rules.
Such objects are generic for nonlinear dissipative dynamics and
hence can have relevance for actual WWW structure. Thus these objects
can find some reflections in the PageRank properties. 
The dynamical system is  described by the Chirikov 
typical map \cite{chirikov}
with dissipation,
the properties of this simple model has been analyzed in detail 
in a recent work \cite{frahm}.
We find that the Google matrix generated by this
dynamical model has many $\lambda_i$ close to $1$ and 
the PageRank becomes sensitive to $\alpha$ (see Fig.~\ref{fig1}).
This model captures also other specific properties of WWW Google matrices.
To construct a network of nodes from a continuous 
two-dimensional phase space we divide the space of dynamical 
variables $(x,y)$ on $N=N_x \times N_y$ cells (we use $N_x=N_y$).
Then $N_c$ trajectories are propagated from a cell $j$ on the whole period
of the dynamical map and
the elements $S_{ij}$ are taken to be equal to a relative number $N_i$ 
of trajectories arrived at 
a cell $i$ ($S_{ij}=N_i/N_c$ and $\sum_i S_{ij}=1$).
Thus ${\bf S}$ gives a coarse-grained
approximation of  the Perron-Frobenius operator for the dynamical map.
The Google matrix ${\bf G}$  of size $N$ is constructed from ${\bf S}$
according to Eq.~(\ref{eq1}). We use a sufficiently large values of 
$N_c$ so that the properties of ${\bf G}$ become not sensitive to $N_c$.

Such a discrete approximation of the Perron-Frobenius operator 
is known in dynamical systems as the Ulam method \cite{ulam}.
Indeed, Ulam conjectured that such a matrix approximant correctly describes
the Perron-Frobenius operator of continuous phase space. For  hyperbolic
maps the Ulam conjecture was proven in  \cite{li}.
Various types of more generic one-dimensional maps
have been studied in \cite{tel,kaufmann,froyland2007}.
Further mathematical results have been reported in 
\cite{ding,liverani,froyland2008a,froyland2008b} with extensions 
and prove of convergence for hyperbolic maps in
higher dimensions. However, the studies of more generic two-dimensional
maps remain rather  restricted (see e.g. \cite{froyland1998}) 
and non-systematic. In principle the construction of directed graphs
on the basis of dynamical systems is a known mathematical approach
(see e.g. \cite{osipenko}) but the spectral properties of 
the Google matrix built on such graphs were not studied till now. 

In this paper we show that the Ulam method applied 
to two-dimensional dissipative dynamical maps generates 
a new type of directed networks which we call
the Ulam networks. We present here numerical and analytical studies of
certain properties of the Google matrix of such networks.

The paper is organized as follows: in Section II we give the description
the Chirikov typical map and the way the Ulam network is constructed on 
the basis of this map with the corresponding Google matrix, 
the properties of the map and the network are described here;
in Section III the properties of the eigenvalues and eigenstates
of the Google matrix are analyzed in detail including the delocalization
transition for the PageRank,  fractal Weyl law and the global contraction
properties; the summary of the results is presented in Section IV.

\section{II Ulam networks of dynamical maps}

\subsection{Chirikov typical map}
To construct an Ulam network and a generated by it Google matrix
we use a dynamical two-dimensional dissipative map.  
The dynamical system is described by the Chirikov typical map
introduced in 1969 for a description of continuous chaotic flows
\cite{chirikov}:
\begin{equation}
y_{t+1} =\eta y_{t}+k \sin (x_t+\theta_t) \;, \;\; x_{t+1} = x_t+y_{t+1} \; .
\label{eq2}
\end{equation}
Here the dynamical variables $x,y$ are taken at 
integer moments of time $t$. Also
$x$ has a meaning of phase variable and $y$ is a conjugated
momentum or action. The  phases
$\theta_t=\theta_{t+T}$ are $T$ random phases periodically repeated
along  time $t$. We stress that their $T$ values 
are chosen and fixed once and they are not
changed during the dynamical evolution of $x,y$.
We consider the map in the region of Fig.~1 
($0 \leq x < 2\pi, -\pi \leq y <\pi$)
with the $2\pi$-periodic boundary conditions. 
The parameter $0< \eta \leq 1$ gives the global dissipation. 
The properties of the symplectic map at $\eta=1$
have been studied recently in detail \cite{frahm}.
The dynamics is globally chaotic for $k > k_c \approx 2.5/T^{3/2}$ and
the Kolmogorov-Sinai entropy is $h \approx 0.29 k^{2/3}$
(more details about chaotic dynamics and
the Kolmogorov-Sinai entropy can be found in
\cite{sinai,mbrin,chirikov79,ott}).

In this study we use two random sets of 
phases $\theta_t$ with $T=10$ and $T=20$.
Their  values are given in the Appendix.
We also fixed the dissipation parameter $\eta=0.99$ for $T=10$
and $\eta=0.97$ for $T=20$. We call these two sets of parameters
as $T10$ and $T20$ sets respectively.
The majority of data are obtained at
$k=0.22$ for the set $T10$ and  at $k=0.3$ for the set $T20$
(see Fig.~\ref{fig1}). These are two main working points 
for this work.

For the set $T10$ ($k=0.22$, $\eta=0.99$)
we have the theoretical value of the Kolmogorov-Sinai
entropy $h=0.29k^{2/3}=0.105$
for the symplectic map at $\eta=1$ \cite{frahm}.
The actual value at $\eta=1$ is determined numerically 
by the computation of the Lyapunov exponent and has a value $h=0.0851$.
For $\eta=0.99$ we also have the global dissipation rate
$\gamma_c=-T\ln \eta=0.1005$ after the map period
(which is equal to $T$ iterations). The global contraction
factor is $\Gamma_c=\eta^T=\exp(-\gamma_c)=0.9043$. 
For a weak dissipation the fractal dimension $d$ of the limiting
set can be approximately estimated 
in a usual way (see e.g. \cite{ott}) as $d=2-\gamma_c/(Th)=1.882$.
%the theoretical value of the
%fractal exponent for data in Fig.~\ref{fig5} inset
%is $\nu=1-\gamma_c/(Th)=1-0.118=0.882$ while the numerical
%fit gives $\nu=0.85$.

\begin{figure}
\centerline{\epsfxsize=8.5cm\epsffile{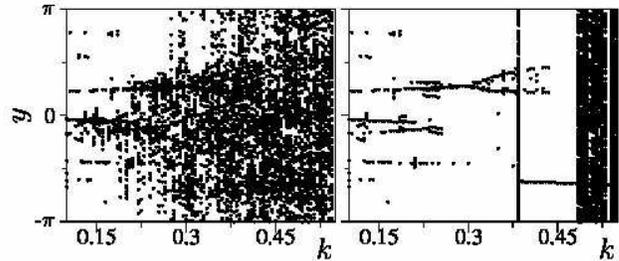}}
\vglue -0.2cm
\caption{Bifurcation diagram
showing values of $y$ vs. map parameter $k$
for the set $T10$ of the Chirikov typical map (\ref{eq2}).
The values of $y$, obtained from 10 trajectories
with initial random positions in the phase space region,
are shown for integer moments of time
$100 < t/T \leq 110$ (left panel) and $10^4 < t/T \leq 10^4+100$ 
(right panel).
}
\label{fig2}
\end{figure}

In a similar way 
for the set $T20$ ($k=0.3$, $\eta=0.97$) we have
the theoretical value
$h=0.29k^{2/3}=0.1299$, while the actual numerical value is $h=0.1081$.
Also here $\gamma_c=-T\ln \eta=0.609$, $\Gamma_c=0.5437$ and 
the estimated fractal dimension of the limiting set is 
$d=2-\gamma_c/(Th)=1.718$.
% the theoretical fractal exponent
%for data in Fig.~\ref{fig5} inset is
%$\nu=1-\gamma_c/(Th)=1-0.282=0.718$ while the numerical
%fit gives $\nu=0.61$.

\begin{figure}
\centerline{\epsfxsize=8.5cm\epsffile{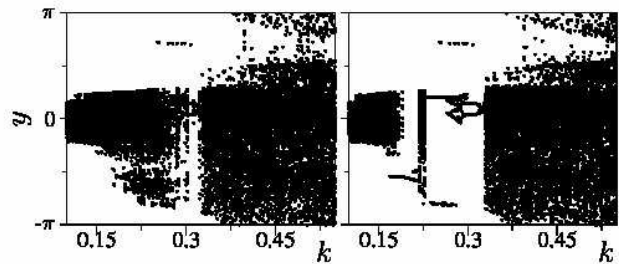}}
\vglue -0.2cm
\caption{Same as in Fig.~\ref{fig2}
for the set $T20$.
}
\label{fig3}
\end{figure}

The bifurcation diagrams for the sets $T10$ and $T20$
are shown in Fig.~\ref{fig2} and Fig.~\ref{fig3} respectively.
On large time scales we clearly see 
parameter $k$ regions with simple and chaotic attractors.
For a shorter time scales a distinction between two regimes
becomes less pronounced. This means that during a long time
a trajectory moves between few simple attractors
(which are clearly seen in Fig.~\ref{fig1} in the left column)
before a final convergence is reached.

\subsection{Network construction and distribution of links}

The Ulam network for the Chirikov typical map (\ref{eq2})
is constructed in the following way. The whole phase space region
$2\pi \times 2\pi$ is divided into $N=N_x \times N_y$ cells
($N_x=N_y$)
and $N_c$ trajectories are propagated from each given cell $j$ during
$T$ map iterations which form the period of the map.
After that the elements of matrix $S_{ij}$ are computed
as $S_{ij}=N_i/N_c(j)$ where $N_i$ is a number of trajectories 
arrived from a cell $j$ to cell $i$. In this way we have by
a definition $\sum_i S_{ij}=1$. Such ${\bf S}$ gives a coarse-grained
approximation of  the Perron-Frobenius operator for the map (\ref{eq2}).
The Google matrix ${\bf G}$  of size $N$ is constructed from ${\bf S}$
according to Eq.~(\ref{eq1}). To construct $S_{ij}$
we usually use $N_c=10^4$ but the properties of ${\bf S}$ are not affected 
by a variation of $N_c$ in the interval $10^3 \leq N_c \leq 10^5$.
Since the cell size is very small it is unimportant
in what way $N_c$ trajectories are distributed inside the cell.
Up to statistical fluctuations,
the values of $S_{ij}$ remains 
the same for homogeneous or random distribution
of $N_c$ trajectories inside a cell. 

Up to $N=22500$ we used exact diagonalization of ${\bf G}$
to determine all eigenvalues $\lambda_i$ and right eigenvectors
$\psi_i$, for larger $N$ up to $N=1.44 \cdot 10^6$ we used
the PRA to determine the PageRank vector. The majority of data
are presented for two typical sets $T10, T20$ of parameters of 
the map (\ref{eq2}) and the PageRanks for various values of $\alpha$
are shown in Fig.~\ref{fig1}. 
For these sets the dynamics has a few
fixed point attractors but it takes a long time  $t \sim 10^3$
to reach them. During this time a trajectory visits various
regions of  phase space. 

It is important to note that the discreteness of  phase space 
linked to a finite cell size produces an important physical
effect which is absent in the original continuous map (\ref{eq2}): 
effectively it introduces an
additional  noise which amplitude $\sigma$
is approximately $\sigma \sim 2\pi/\sqrt{N}$.
This becomes especially clear for the symplectic case at $\eta=1$
and at small values of $k$ at $T=1$ (all $\theta_t$ are the same).
In this case the map is reduced to the Chirikov standard map 
\cite{chirikov79} and the continuous map dynamics is bounded
by the invariant Kolmogorov-Arnold-Moser (KAM) curves.
However, the discreteness of  phase space allows to
jump from one cell to another and thus to jump
from one curve to another. This leads to a diffusion in
$y$ direction and appearance of a homogeneous ergodic state 
at $\lambda=1$. A direct analysis
also shows that at any finite cell size the operator ${\bf S}$
has a homogeneous ergodic state with $\lambda=1$, we also
checked this via numerical diagonalization of matrix sizes
$N \approx 20000$. This example shows that the Ulam
conjecture is not valid for quasi-integrable  symplectic maps
in the KAM regime. 

\begin{figure}
\centerline{\epsfxsize=8.5cm\epsffile{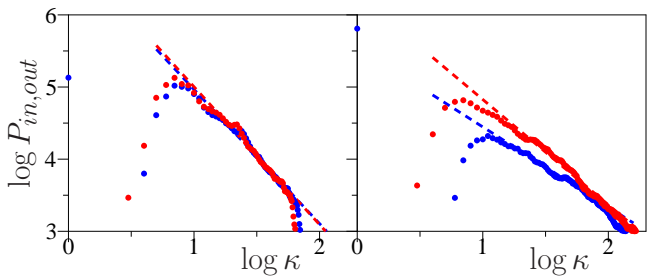}}
\vglue -0.2cm
\caption{(Color online) Differential distribution of number of nodes
with
{\it ingoing} $P_{in}(\kappa)$ (blue) and
{\it outgoing} $P_{out}(\kappa)$ (red) links $\kappa$
for sets $T10$ (left) and $T20$ (right).
The straight dashed lines give the algebraic fit
$P(\kappa) \sim \kappa^{-\mu}$
with the exponent $\mu = 1.86, 1.11$ ($T10, T20$) for {\it ingoing} 
 and $\mu = 1.91, 1.46$ ($T10, T20$)
{\it outgoing}  links.
Here $N=1.44 \cdot 10^6$ and $P(\kappa)$ gives a number
of nodes at a given integer number of links $\kappa$ for this matrix size.
Blue point at $\kappa=0$ shows that in the whole matrix there is 
a significant number of nodes with zero {\it ingoing} links.
}
\label{fig4}
\end{figure}

\begin{figure}
\centerline{\epsfxsize=8.5cm\epsffile{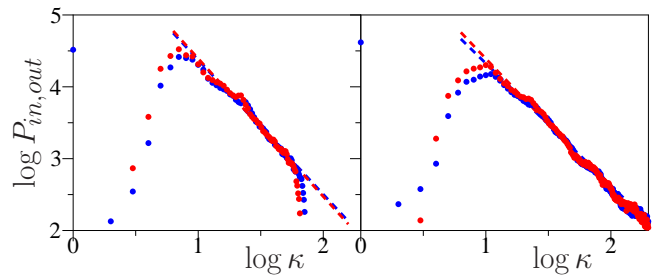}}
\vglue -0.2cm
\caption{(Color online) Same as in Fig.~\ref{fig3}
for the set $T10$ at $k=0.22$ (left) (same as Fig.~\ref{fig4} left)
and $k=0.6$ (right) and $N=3.6 \cdot 10^5$.
The fit gives the exponent $\mu=1.87, 1.92$
for {\it ingoing} (blue), {\it outgoing} (red) links at $k=0.22$ (left)
and  
$\mu=1.70, 1.83$
for {\it ingoing} (blue), {\it outgoing} (red) links at $k=0.6$ (right).
}
\label{fig5}
\end{figure}

The physical origin of the difference between 
the continuous map and the finite size cell approximation
is due to introduction of an effective noise term $\sigma_t$
in r.h.s. of (\ref{eq2}) induced by a finite cell size.
Due to this noise the trajectories diffuse over all 
region $-\pi <y<\pi$ after a 
diffusive time scale $t_D \sim \pi^2/\sigma^2$ even
if the continuous map is in the KAM regime with bounded dynamics in $y$.
Hence, here $\sigma \sim 2\pi/\sqrt{N}$ is an effective amplitude of noise
introduced by cell discreetness. 

Even if this $\sigma$-noise leads to a drastic change
of dynamics for quasi-integrable regime its effects are 
not very important in the case of chaotic dynamics
where noise gives only a small additional variation
as compared to strong dynamical variations induced by dynamical chaos.
With such a physical understanding of discreetness effects
we continue to investigate the properties of 
the Ulam networks. However, we stress that the $\sigma$-noise
is local in the phase space and hence it is qualitatively different
from the Google term $\alpha$ which generates stochastic jumps
over all sites.

In Figs.~\ref{fig4},\ref{fig5} we show
the distributions of 
ingoing $P_{in}(\kappa)$ and outgoing  $P_{out}(\kappa)$ links $\kappa$
in the Ulam network presented by ${\bf S}$ 
matrix generated by the map (\ref{eq2}) as described above.
These distributions are satisfactory described by a scale-free algebraic decay
$P \sim 1/\kappa^{\mu}$ with 
$\mu \approx 1.86, 1.11$ for
ingoing and $1.91, 1.46$ outgoing 
links at $T10,T20$ respectively and a typical number of links per
node $\kappa \sim 10$
(see Fig.~\ref{fig4} and Fig.~\ref{fig5}).
Such values are compatible with the WWW data of scale-free type
where $\mu \approx 2.1, 2.7$
for ingoing, outgoing links \cite{googlebook,donato}.
However, we may also note an appearance of certain deviations 
at large values of $\kappa$. Indeed, for a dynamical system
a large number of links appears due to exponential
stretching of one cell after $T$ map iterations
that gives a typical number of links $k \sim \exp(hT)$.
It is possible that during the dynamical evolution much larger
values of stretching can appear. Indeed, the comparison of 
two cases at $k=0.22$ and $k=0.6$ for the set $T10$ in Fig.~\ref{fig5}
shows that for larger $k$ the scale-free distribution continues
to much larger values of $\kappa > 200$ while for smaller $k$
the  scale-free type decay stops around $\kappa \approx 50$.
For the set $T20$ the stretching is stronger and the scale-free 
decay continues up to larger values of $\kappa$.

It is clear that for the Ulam networks discussed here
one has a rapid exponential decay of links distribution
at asymptotically large link number $\kappa$.
However, due to an exponential growth of typical 
$\kappa \sim \exp(hT)$ a scale-free type decay
can be realized up to very large $\kappa$ by
increasing $T$. In these studies we stay at the
chosen working points where a scale-free decay
remains dominant for matrix sizes of the 
order of $N \sim 10^5 - 10^6$.

Finally we note that the models of the Google matrix generated by
the Ulam networks are most interesting for dissipative maps. 
Indeed, by construction the left eigenvector
of the Google matrix  $\psi_i^+ {\bf G} = \psi_i^+$ at $\lambda=1$
is a homogeneous vector $\psi_i^+=const$. As a result
for symplectic maps the right vector of PageRank $p_j$
is also homogeneous. Only dissipation term generates 
an inhomogeneous decay of $p_j$.

\section{III Properties of eigenvalues and eigenstates}

\subsection{Delocalization transition for PageRank with $\alpha$}

The variation of PageRank $p_j$ with $\alpha$ is shown in Fig.~\ref{fig1}
for two sets $T10$ and $T20$. The distribution $p_j$
is plotted for each cell of the phase space $(x,y)$,
the numbering of cells is done by the integer grid
$n_x \times n_y$ which has a certain correspondence
with the index $j$ which numerates the values of $p_j$
in the decreasing order with $j$. 
At $\alpha=1$ 
the distribution $p_j$ is concentrated only on a few local spots
corresponding to fixed point attractors. 
Physically this happens due to presence of $\sigma$ noise, induced by 
cell discretization,  which leads to transitions between various 
fixed points. With the decrease of $\alpha$
the PageRank starts to spread over a strange attractor set. 
The properties of strange attractors in dynamical dissipative
systems are described in \cite{ott}.
In the map (\ref{eq2}) the strange attractor appears at larger values
of $k$ (namely $k > 0.5$ for $T10$, $k>0.34$ for $T20$,
see Figs.~\ref{fig2},~\ref{fig3})
but a presence of effective noise induced by $\sigma$ and $1-\alpha$
terms leads to an earlier emergence of strange attractor.
Below a certain value  $\alpha < \alpha_c$ the PageRank becomes
completely delocalized over the strange attractor as it is clearly
seen in Fig.~\ref{fig1} for the set $T10$.

\begin{figure}
\centerline{\epsfxsize=8.5cm\epsffile{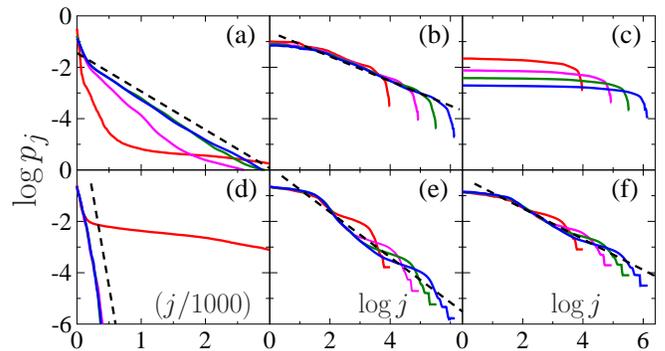}}
\vglue -0.2cm
\caption{(Color online) PageRank distribution $p_j$ 
for $N=10^4$, $9\cdot 10^4$, $3.6\cdot 10^5$ and $1.44\cdot 10^6$ 
shown by red, magenta, green and blue curves, 
the dashed straight lines show fits 
$p_j \sim 1/j^\beta$ with $\beta$:
 $0.48$ (b), $0.88$ (e), 
$0.60$ (f). Dashed lines in panels (a),(d)
show an exponential Boltzmann decay (see text, lines are shifted
in $j$ for clarity).
Other parameters, including the values of $\alpha$, and panel order 
are as in Fig.~\ref{fig1}. 
In panels (a),(d) the curves at large $N$ become superimposed.
Here and below logarithms are decimal.
%dashed fits $p_j=a\exp(-bj)$
%$a=\exp(-4.01); b=3.44 \cdot 10^{-3}$ for panel (a)
%$a=\exp(-0.55); b=0.033$ for panel (d)
}
\label{fig6}
\end{figure}

The dependence of $p_j$ on $j$ 
is shown in more detail in Fig.~\ref{fig6}.
For $\alpha=1$ PageRank shows a rapid drop with $j$
that can be fitted by an exponential Boltzmann type distribution
$p_j \sim \exp(- b \gamma_c j/D_\sigma)$ where $b$ is a numerical constant
($b \approx 1.4; 2.1$ for $T10; T20$), $\gamma_c= -T \ln \eta$
is the global dissipation rate and $D_\sigma = \sigma^2 N \approx (2\pi)^2$
is $\sigma$ noise diffusion
(dashed lines in Fig.~\ref{fig6}a,d).
Such an exponential decay results from the Fokker-Planck
description of map (\ref{eq2}) in the presence of $\sigma$ noise
term which gives diffusive transitions on nearby cells.
For $\alpha <1$ random surfer transitions introduced by Google
give a significant modification of PageRank which
shows an algebraic decay $p_j \sim 1/j^{\beta}$ with 
the exponent $\beta$ dependent on $\alpha$ (Fig.~\ref{fig6}b,e,f);
for the set $T20$ at $\alpha=0.95$ we obtain $\beta \approx 0.88$
being close to the numerical value found for the WWW \cite{googlebook}.
However, $\beta$ decreases with the decrease of $\alpha$ and for 
$T10$ set a delocalization takes place for $\alpha=0.85$
so that $p_j$ spreads homogeneously over the strange attractor
(see Fig.~\ref{fig1} top right panel and Fig.~\ref{fig6}c).
 For $T20$ set $p_j \sim \psi_{i=1}(j)$ remains localized
at $\alpha=0.85$ so that a PArticipation Ratio (PAR)
$\xi=\sum_j (|\psi_i(j)|^2)^2/\sum_j(|\psi_i(j)|^4$ for the PageRank
remains finite at large $N$. We use this definion of PAR $\xi$
for all eigenvectors $\psi_i(j)$.

\subsection{Properties of other eigenvectors}

\begin{figure}
\centerline{\epsfxsize=8.5cm\epsffile{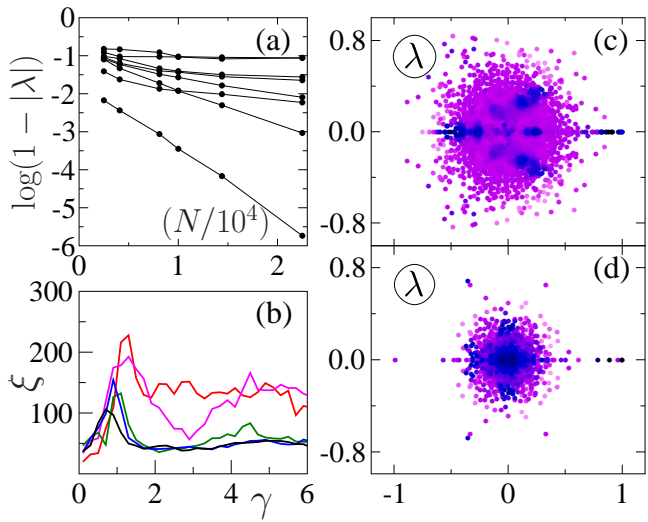}}
\vglue -0.2cm
\caption{(Color online) 
(a) Dependence of gap $1-|\lambda|$ on Google matrix size $N$
for few eigenstates with $|\lambda|$ most close to $1$, set $T10$, $\alpha=1$;
(b) dependence of PAR $\xi$ on $\gamma=-2 \ln |\lambda|$
for $N=2500$, $5625$, $8100$, $10^4$, $14400$
for set $T10$, $\alpha=1$ (curves from top to bottom: 
red, magenta, green, blue, black);
(c) complex plane of eigenvalues $\lambda$ for set $T10$
with their PAR $\xi$ values shown by grayness (black/blue for minimal
$\xi \approx 4$, gray/light magenta for maximal $\xi \approx 300$;
here  $\alpha=1$, $N=1.44 \cdot 10^4$);
(d) same as (c) but for set $T20$.
}
\label{fig7}
\end{figure}

To understand the origin of the delocalization transition in $\alpha$
we analyze  in Fig.~\ref{fig7} the properties of all eigenvalues 
$\lambda_i$
and eigenvectors $\psi_i$ with their PAR $\xi$.  
Due to $\sigma$ noise activation
transitions take place between the attractor fixed points leading to
states with $\lambda_i$ being exponentially close to
$\lambda=1$ (Fig.~\ref{fig7}a). 
The convergence to $|\lambda|=1$ is exponential in $N$
for certain states and may lead to numerical problems
at very large $N$. However, the standard
numerical diagonalization methods remained stable for the
values of $N$ used in our studies.

The distribution of $\lambda_i$
in the complex plane is shown in Fig.~\ref{fig7}c,d: there are
$\lambda_i$ approaching $\lambda=1$ mainly along the real axis
but a majority of $\lambda_i$ are distributed inside a circle
of finite radius around $\lambda=0$; this radius
decreases with the increase of global dissipation from
$\gamma_c=0.10$ for set $T10$ to $\gamma_c=0.61$ for $T20$.
The PAR values for states inside the circle
have typical values $4 \leq \xi \leq 300$ shown by grayness.
The dependence of $\xi$ on $\gamma=-2 \ln |\lambda|$ 
and $N$ shows that
the eigenstates inside the circle remain localized at
large $N$ (Fig.~\ref{fig7}b). We attribute this to the fact that
at large $N$ the diffusion due to $\sigma$ noise 
in presence of dissipation leads to spreading only over 
a finite number of cells and thus $\xi$ remains bounded.
This $\xi(\gamma,N)$ dependence is different from 
one obtained in \cite{ggs} for the Albert-Barabasi model,
the comparison with data from WWW University networks is 
less conclusive due to strong fluctuations from
one network to another (see Fig.~4 in \cite{ggs}):
an average growth of $\xi$ is visible there even if
at $N \sim 10^4$ the values of $\xi$
are comparable with those of Fig.~\ref{fig7}b.
Globally our data of Fig.~\ref{fig7} show that the 
diffusive modes at $|\lambda_i| < 1$ remain localized on
a number of nodes $\xi \ll N$. 

We also stress an important property of 
eigenvalues and eigenvectors with $0<|\lambda_i| < 1$.
In agreement with
the known theorems \cite{googlebook}
our numerical data show that for the states
with  $0<|\lambda_i| < 1$ their $\xi_i$  are independent of $\alpha$
($\lambda_i$ are simply rescaled by a factor $(1-\alpha)$ 
according to \cite{googlebook}).
This happens due to a specific property of $(1-\alpha){\bf E}/N$ term
in ${\bf G}$, which is constructed from a homogeneous vector with rank
equal to unity. Right eigenvectors are orthogonal to the homogeneous
left vector and hence $(1-\alpha)$ term affects only the PageRank but 
not other eigenvectors. 

\subsection{Fractal Weyl law for Google matrix}

\begin{figure}
\centerline{\epsfxsize=8.5cm\epsffile{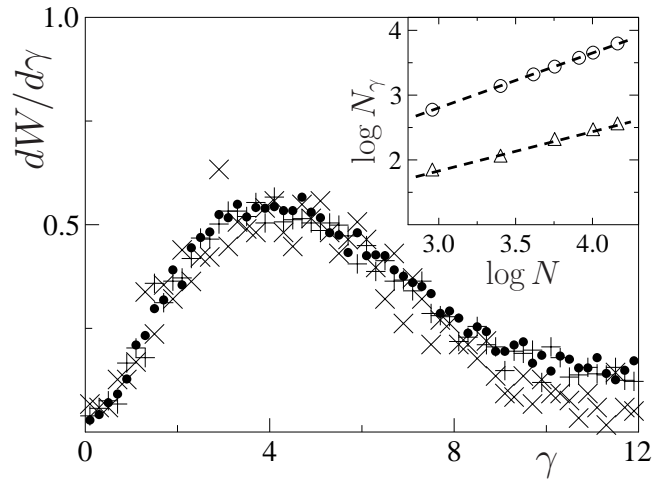}}
\vglue -0.2cm
\caption{Probability distribution $d W(\gamma)/d\gamma$
for set $T10$, $\alpha=1$ at $N=2.5 \cdot 10^3 (\times)$,
$10^4 (+)$, $1.44 \cdot 10^4$ (dots); $W(\gamma)$ is normalized by the
number of states $N_\gamma=0.55 N^{0.85}$ with $\gamma <6$.
Inset: dependence of number of states $N_\gamma$ with $\gamma<\gamma_b$
on $N$ for sets $T10$ (circles, $\gamma_b=6$) and $T20$ 
(triangles, $\gamma_b=3$);
dashed lines show the fit
$N_\gamma=A N^{\nu}$ with
$A=0.55, \nu= 0.85$ 
%(theory value $\nu_{th}=0.88$)  
and 
$A=0.97, \nu =0.61$  
%($\nu_{th} = 0.72$) 
respectively.
}
\label{fig8}
\end{figure}

Another interesting characteristic of ${\bf G}$ is
the density distribution $d W(\gamma)/d \gamma$ over $\gamma$.
The data presented in  Fig.~\ref{fig8} show that its form becomes
size independent in the limit of large $N$.
At small $\gamma < 3$ the density decreases approximately 
linearly with $\gamma$ without any large gap.
We find rather interesting that the total
number of states $N_\gamma$ with finite 
$\gamma < \gamma_b \approx 5$ grows 
algebraically as $N_\gamma = A N^{\nu}$
with $\nu < 1$ (Fig.~\ref{fig8} inset).
We interpret this result on the basis of the fractal Weyl law
established recently for non-unitary matrices with fractal eigenstates
(see e.g. \cite{zworski,dlsweyl} and Refs. therein). According to this law
the exponent $\nu$ is $\nu=d-1$ where $d$ is the fractal dimension
of the system. Approximately we have $d-1 \approx 1 - \gamma_c/(T h)$
\cite{ott,dlsweyl} that gives $\nu=0.88, 0.72$ for the sets $T10, T20$ with 
the numerical values of $\gamma_c$, $h$ given above. These values are
in a good agreement with the fit data $\nu=0.85, 0.61$ 
of Fig.~\ref{fig8} inset.
The fact that $\nu <1$ implies that almost all states
have $\lambda=0$ in the limit of large $N$
(in this work we do not discuss the properties 
of these degenerate states with large $\xi \sim N$).

It is interesting to note that the fractal Weyl law is usually discussed
for the open quantum chaos systems 
(see \cite{zworski,dlsweyl} and Refs. therein). There the matrix size
is inversely proportional to an effective Planck constant $N \propto 1/\hbar$.
For the Ulam networks generated by dynamical attractors 
a cell size in the phase space places the role of
effective $\hbar$. This opens interesting parallels
between quantum chaotic scattering and discrete matrix representation
of the Perron-Frobenius operators of dynamical systems. 

\subsection{PageRank delocalization again}

\begin{figure}
\centerline{\epsfxsize=8.5cm\epsffile{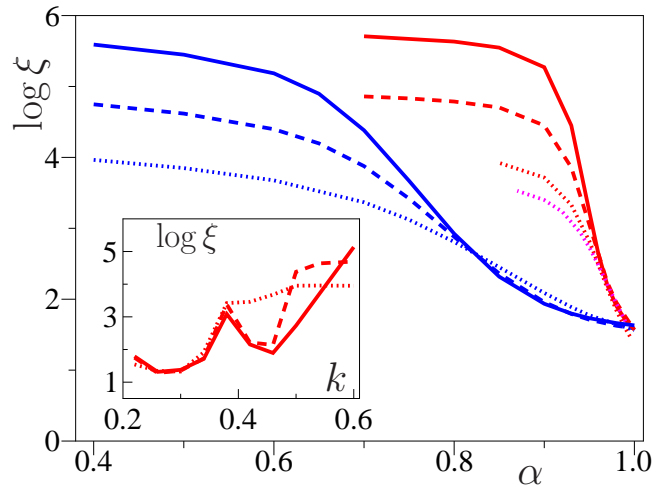}}
\vglue -0.2cm
\caption{(Color online) Dependence of PageRank $\xi$
on $\alpha$
for set $T10$ at $N=5625$ (dotted magenta),
$1.44 \cdot 10^4$ (dotted red), $9 \cdot 10^4$ (dashed red),
$6.4 \cdot 10^5$ (full red) and for $T20$
at $N=1.44 \cdot 10^4$ (dotted blue), $9 \cdot 10^4$ (dashed blue),
$6.4 \cdot 10^5$ (full blue). Inset shows dependence of $\xi$
on $k$ for set $T10$ at $\alpha=0.99$ with 
$N=1.44 \cdot 10^4$ (dotted red), $9 \cdot 10^4$ (dashed red),
$3.6 \cdot 10^5$ (full red).
}
\label{fig9}
\end{figure}

\begin{figure}
\centerline{\epsfxsize=4.2cm\epsffile{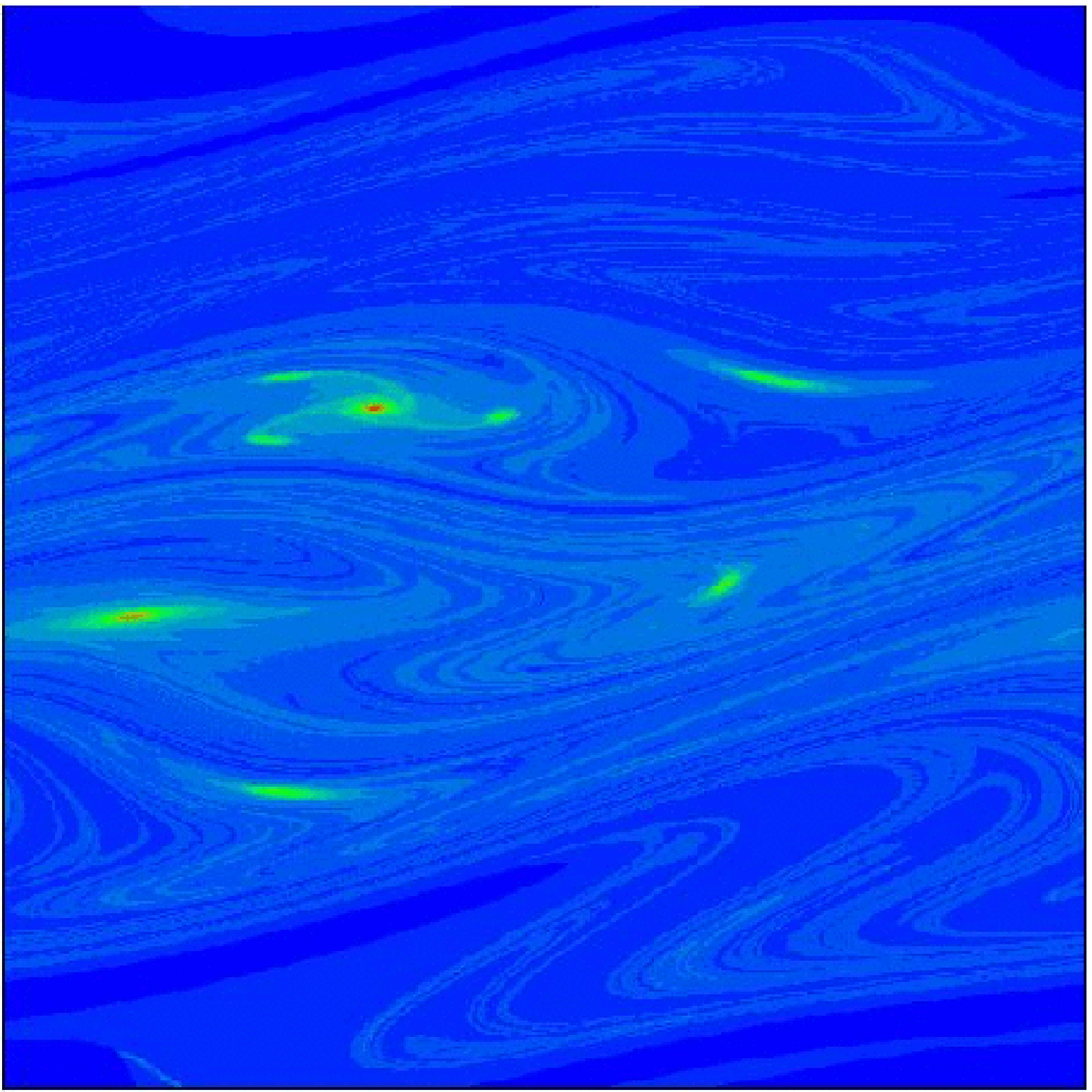}
\hfill\epsfxsize=4.2cm\epsffile{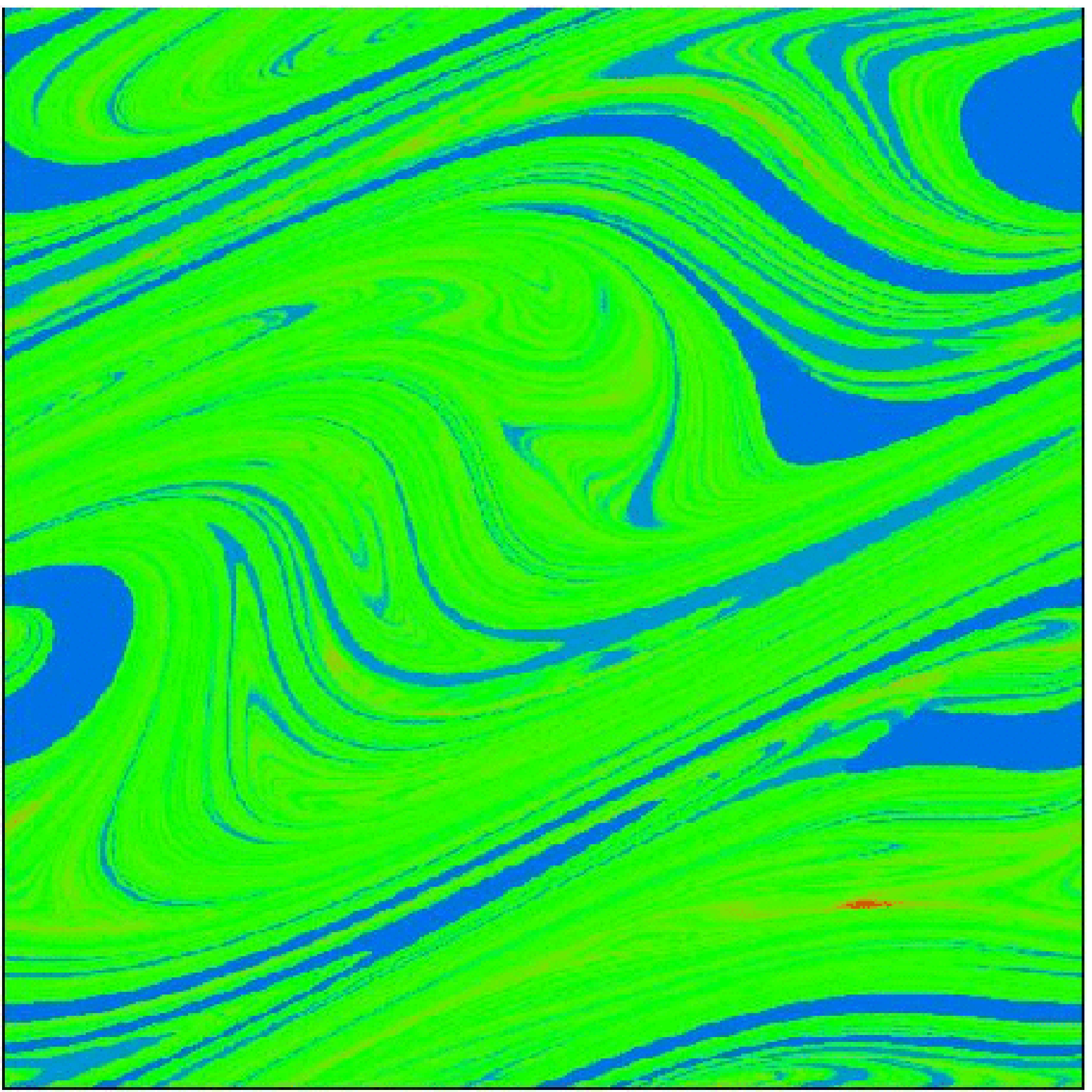}}
\vglue -0.2cm
\caption{(Color online) Same as Fig.~\ref{fig1}
for the set $T10$ at $\alpha=0.99$, $N=3.6 \cdot 10^5$
at $k=0.22$ (left) and $k=0.6$ (right); PAR $\xi$
are the same as in the inset of Fig.~\ref{fig9}.
}
\label{fig10}
\end{figure}

The dependence of PAR $\xi$ of the PageRank on
$\alpha$ and $N$ is shown in Fig.~\ref{fig9}. It permits to determine 
the critical value $\alpha_c$ below which the PageRank becomes
delocalized showing $\xi$ growing with $N$.
According to this definition we have $\xi$ independent of large $N$
for $\alpha > \alpha_c$ while for $\alpha < \alpha_c$
the PAR $\xi$ grows with $N$.
The obtained data give $\alpha_c \approx 0.95$, $0.8$ for $T10, T20$. 
Further investigations are needed to understand  
the dependence of $\alpha_c$ on system parameters. Here we make a 
conjecture that $1-\alpha_c \approx C \gamma_c \ll 1$ 
with a numerical constant $C \approx 0.3$.
Indeed, for larger dissipation rate
$\gamma_c = - T \ln \eta$ a radius of 
a circle with large density of $\lambda_i$ in the complex plane $\lambda$
becomes smaller (see Fig.~\ref{fig7}c,d)
and thus larger values of $1-\alpha$
are required to have a significant contribution
of these excited relaxation modes to the PageRank.
Also the data of \cite{dlsweyl} for systems with absorption rate
$\gamma_c$ show a low density of states at $\gamma < \gamma_c$
so that it is natural to expect that one should have 
$1-\alpha_c \sim \gamma_c$ to get a significant contribution
of delocalized relaxation modes from a strange attractor set
to  the PageRank.  It is quite probable that $C$ depends in addition
on system parameters. Indeed, even at fixed $\gamma_c$
and $\alpha=0.99$ being rather close to $1$ it is possible to
have a transition from localized to 
delocalized  PageRank by increasing $k$ in the map (\ref{eq2})
(see Fig.~\ref{fig9} inset and Fig.~\ref{fig10}). 
This transition in $k$ takes place
approximately at $k \approx 0.55$ when fixed point attractors
merge into a strange attractor (see the bifurcation 
diagram in Fig.~\ref{fig2}). 
A peak in $\xi$ around $k \approx 0.38$
is related to birth and disappearance of a strange attractor
in a narrow interval of $k$ at $k \approx 0.38$. 
At the same time an increase of $k$ from $0.22$ to
$0.6$ practically does not affect the link distributions
$P(\kappa)$ changing the value of $\mu$
only by 10\% (see Fig~\ref{fig5}). This shows that the correlations
inside the directed network generated by the map (\ref{eq2})
play a very important role.

\subsection{Global contraction}

As discussed above a nontrivial decay
of the PageRank $p_j$ in our Ulam network 
appears due to a dissipative nature of the map (\ref{eq2}).
Indeed, since $\eta <1$ there is a global contraction of the phase
space area by a factor $\Gamma_c=\eta^T$ after
$T$ iterations of the map (after its period).
Such a property is very natural for the continuous map
but it is more difficult to see its signature
from the matrix form of the Perron-Frobenius operator
after the introduction of discreteness of the phase space.

Nevertheless this contraction can be extracted from the matrix
${\bf G}$ taken at $\alpha=1$.
To extract it we apply ${\bf G}$ with $\alpha=1$ to a homogeneous vector 
$p^{(h)}_j=1/N$ getting the new vector ${\bar p^{(h)}} = {\bf G} p^{(h)}$
and count the number of nodes $N_\Gamma$
where ${\bar p^{(h)}} > q/N  $ and $0<q<1$ is some positive number
characterizing the level of the distribution.
Then the contraction of the network is defined 
as a fraction of such states: $\Gamma=N_\Gamma/N$.

\begin{figure}
\centerline{\epsfxsize=4.2cm\epsffile{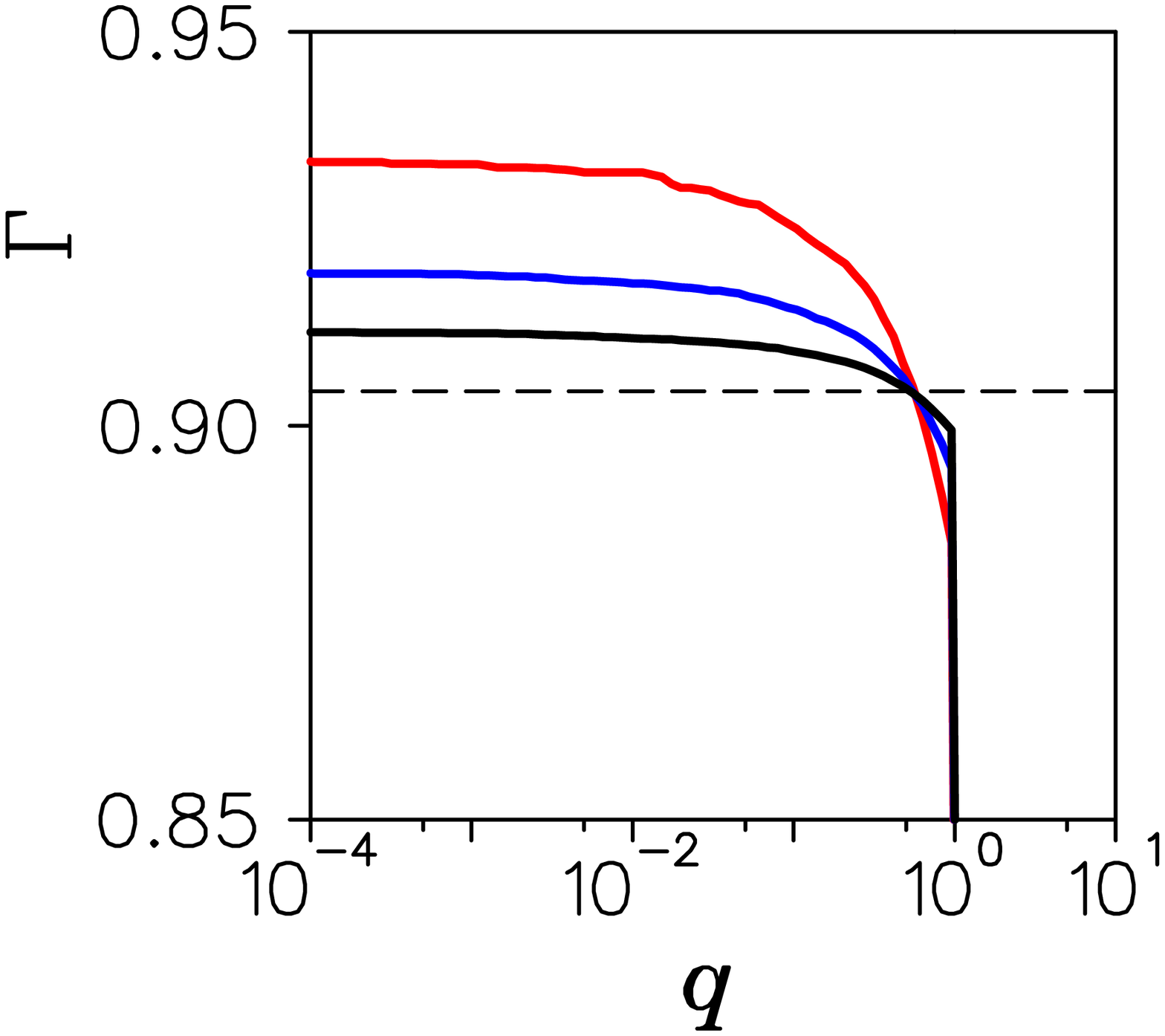}
\hfill\epsfxsize=4.2cm\epsffile{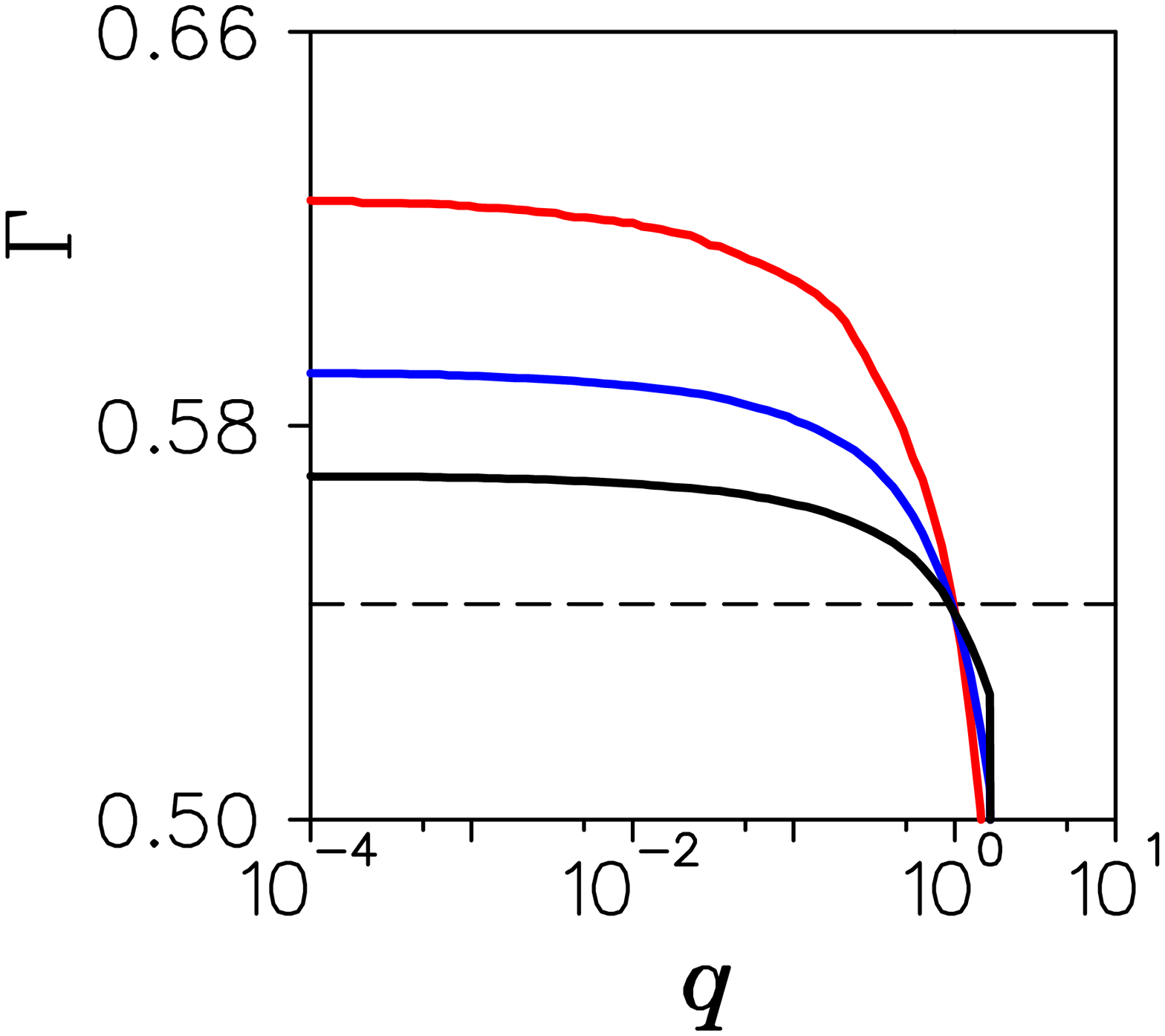}}
\vglue -0.2cm
\caption{(Color online) Dependence of the network contraction
factor $\Gamma$ on the level $q$ of probability distribution
 over the network nodes (see text). 
Left panel shows data for the set $T10$ 
at $k=0.22$, right panel shows data for the set $T20$ at $k=0.3$
for the Ulam network of map (\ref{eq2}).
The size of the network is $N=10^4, 4 \cdot 10^4, 16 \cdot 10^4$
(curves from top to bottom at $q=0.01$). The dashed curves show
the contraction $\Gamma_c=\eta^T$ of the continuous map (\ref{eq2})
corresponding to the network with $N=\infty$.
}
\label{fig11}
\end{figure}

The result of computation of the contraction factor for the Ulam network
of map (\ref{eq2}) for the sets $T10$, $T20$ is shown in Fig.~\ref{fig11}.
The network contraction parameter $\Gamma$ is independent of $q$
in a large interval $10^{-4} \leq q \leq 0.1$
and it converges to the contraction value $\Gamma_c$
of a continuous map in the limit of large matrix size $N$.

We think that the Google matrix of WWW networks can 
be also characterized by a global contraction factor
and it would be interesting to study its properties
in more detail. However, this remains a task for future studies.

\section{IV Summary}

In summary, we demonstrated that the Perron-Frobenius 
operator built from a simple dissipative map
with dynamical attractors generates a scale-free
directed network with properties being rather similar to
the WWW. The networks and their Google matrices are obtained 
on the basis of the Ulam method for coarse-graining
of the Perron-Frobenius operator and thus can be viewed
as the Ulam networks or Ulam graphs.
The Google matrix of such Ulam networks
reproduces many properties of real networks
with an algebraic decay of the PageRank and
quasi-degeneracy of eigenvalues near unity
for the Google parameter  $\alpha=1$.
In this formulation the popular websites correspond to
dynamical fixed point attractors which help to generate
global scale-free properties of the network. 
The PageRank of the system becomes delocalized
for $\alpha$ smaller than a certain critical value,
such a delocalization is linked to emergence of a strange
attractor.  Even for $\alpha$ very close to unity
a moderate change of system parameters
can drive the system to a strange attractor regime
with a complete delocalization of the PageRank
making the Google search inefficient. In view of a great importance
of the Google search for WWW \cite{googlebook,donato}
and its  new emerging applications \cite{redner} it may be
rather useful to study in more detail
the properties of the Google matrix
generated by simple dynamical maps.

Of course, it is quite possible that at the present state 
the Google matrix of WWW is more stable in respect to
variation of $\alpha$ (indications
for that can be found e.g. in \cite{avrach1,avrach2,avrach3}).
However, WWW evolves with time and may become more
sensitive to changes of $\alpha$. Also the Google search 
can be applied to a large variety of other important networks
(see e.g. \cite{avrach3,redner}) which 
may be more sensitive to various parameter variations.
It is quite possible that the Ulam networks discussed here
only partially simulate the properties of the WWW. However,
the Ulam networks are easy to generate and at the same
time they show a large variety of rich interesting properties.
The parallels between the Ulam networks and the actual WWW
can be instructive for deeper understanding of both. 
Therefore, we think that their further studies
will give us better understanding of the Google matrix properties.
The studies of the Ulam networks will also lead to a better
understanding of intricate spectral properties of the Perron-Frobenius 
operators. The application of the thermodynamical formalism 
\cite{ruelle,artuso} to the spectra of such operators 
can help to understand their properties in a better way.

\section{Acknowledgements}

We thank A.S.Pikovsky who pointed to us a link between our numerical
construction procedure of the matrix ${\bf S}$ built 
from the discrete phase space cells and the Ulam method.
One of us (DLS) thanks A.S.Pikovsky for useful discussions
and hospitality at the Univ. Potsdam during the work on the revised
version of this paper. We also thank an unknown referee B
who pointed to us Refs.~\cite{osipenko,boldi,avrach1,avrach2,avrach3}
in the report for the initial short version of the paper.

%\newpage
\section{APPENDIX}

%all figs are at
%/h2/zhirov/Work/GM/figs/prl

The Chirikov typical map (\ref{eq2}) is studied here for
the following random phases $\theta_t/2\pi$
for the set $T10$ :

\noindent
$0.562579$, $0.279666$, $0.864585$, $0.654365$, $0.821395$,
$0.981145$, $0.478149$, $0.834115$, $0.180307$, $0.15902$

\noindent 
and for the set $T20$:

\noindent
$0.415733267627$, $0.310795551489$, $0.632094907846$, $0.749488203411$,
$0.924301928270$, $0.635937571045$, $0.118768635110$, $0.647524548037$,
$0.651928927275$, $0.952312529146$, $0.370553510280$, $0.810837257644$,
$0.814808044380$, $0.834758628241$, $0.993694010264$, $0.702057578688$,
$0.828693568678$, $0.855421638697$, $0.278538720979$, $0.653773338142$.

The numbers are ordered in the serpentine order for $t=1,2,...T$.

After each $T$ iterations the values of $y$ are reduced
inside the interval $(-\pi,\pi)$ corresponding to 
the periodic boundary conditions.

\end{document}